\documentclass{jpsj-suppl}
\usepackage{txfonts} 

\newcommand{\apj}{ApJ}

\title{PUSHing Core-Collapse Supernovae to Explosions in Spherical Symmetry: Nucleosynthesis Yields}

\author{Sanjana \textsc{Sinha}$^{1}$, Carla \textsc{Fr\"ohlich}$^{1}$, Kevin \textsc{Ebinger}$^{2}$, Albino \textsc{Perego}$^{3}$, Matthias \textsc{Hempel}$^{2}$, Marius \textsc{Eichler}$^{2, 3}$, Matthias \textsc{Liebend\"orfer}$^{2}$, and Friedrich-Karl \textsc{Thielemann}$^{2}$}

\inst{$^{1}$Department of Physics, North Carolina State University, Raleigh, NC, 29695-8202, USA \\
$^{2}$Department f\"ur Physik, Universit\"at Basel, CH-4056 Basel, Switzerland \\
$^{3}$Institut f\"ur Kernphysik, Technische Universit\"at Darmstadt, D-64289 Darmstadt, Germany}

\email{ssanjan@ncsu.edu}

\recdate{Aug 20, 2016}

\abst{Core-collapse supernovae (CCSNe) are the extremely energetic deaths of massive stars. They play a vital role in the synthesis and dissemination of many heavy elements in the universe. In the past, CCSN nucleosynthesis calculations have relied on artificial explosion methods that do not adequately capture the physics of the innermost layers of the star. The PUSH method, calibrated against SN1987A, utilizes the energy of heavy-flavor neutrinos emitted by the proto-neutron star (PNS) to trigger parametrized explosions. This makes it possible to follow the consistent evolution of the PNS and to ensure a more accurate treatment of the electron fraction of the ejecta. Here, we present the Iron group nucleosynthesis results for core-collapse supernovae, exploded with PUSH, for two different progenitor series. Comparisons of the calculated yields to observational metal-poor star data are also presented. Nucleosynthesis yields will be calculated for all elements and over a wide range of progenitor masses. These yields can be immensely useful for models of galactic chemical evolution.}

\kword{stars: supernovae: general, nucleosynthesis}

\begin{document}
\maketitle

\section{Introduction}

The detailed mechanism of core-collapse supernovae is still an open question which is being investigated using sophisticated multi-dimensional models. However, at the present time, these models are too computationally expensive for systematic nucleosynthesis studies of multiple progenitors. For a recent review of multi-dimensional core-collapse simulations, see Janka et al. (2016)\cite{janka} and references therein. In order to make nucleosynthesis predictions and discover general trends, we still require robust and readily calculable models which capture the relevant physics of the explosion.

The PUSH method \cite{push} triggers explosions in otherwise non-exploding simulations by parametrically increasing the efficiency of neutrino energy deposition inside the gain region. This is done by depositing a fraction of the luminosity of heavy-flavor neutrinos (emitted by the PNS) behind the shock. The mass cut emerges naturally in the simulation and the electron fraction is followed consistently. Despite their effective character, PUSH models are robust, computationally affordable and self-consistent. We present the Iron group nucleosynthesis yields for CCSNe, calculated for models from two progenitor series, exploded using PUSH.

\section{Methods}

The evolution of the electron fraction ($Y_e$) strongly affects nucleosynthesis in the innermost ejected stellar layers, where the predominant production of Iron group elements occurs. The $Y_e$ can change due to electron neutrino and anti-neutrino interactions. 

PUSH simulations follow the $Y_e$ evolution consistently and provide input trajectories for post-processing, as outlined in \cite{push}. The isotropic diffusion source approximation (IDSA)\cite{idsa} is employed for electron neutrino and anti-neutrino transport while an advanced spectral leakage (ASL)\cite{asl} scheme is used for transport of heavy-lepton flavor neutrinos. The nuclear reaction network, CFNET, follows the abundances of $\sim$ 2000 isotopes to compute the composition of the supernova ejecta. The isotopes included cover the neutron-deficient as well as the neutron-rich side of the valley of $\beta$-stability.

\section{Results}

\begin{figure}[tbh]
\centering
\begin{tabular}{ll}
\hspace{-.015\textwidth}\includegraphics[width=0.48\textwidth]{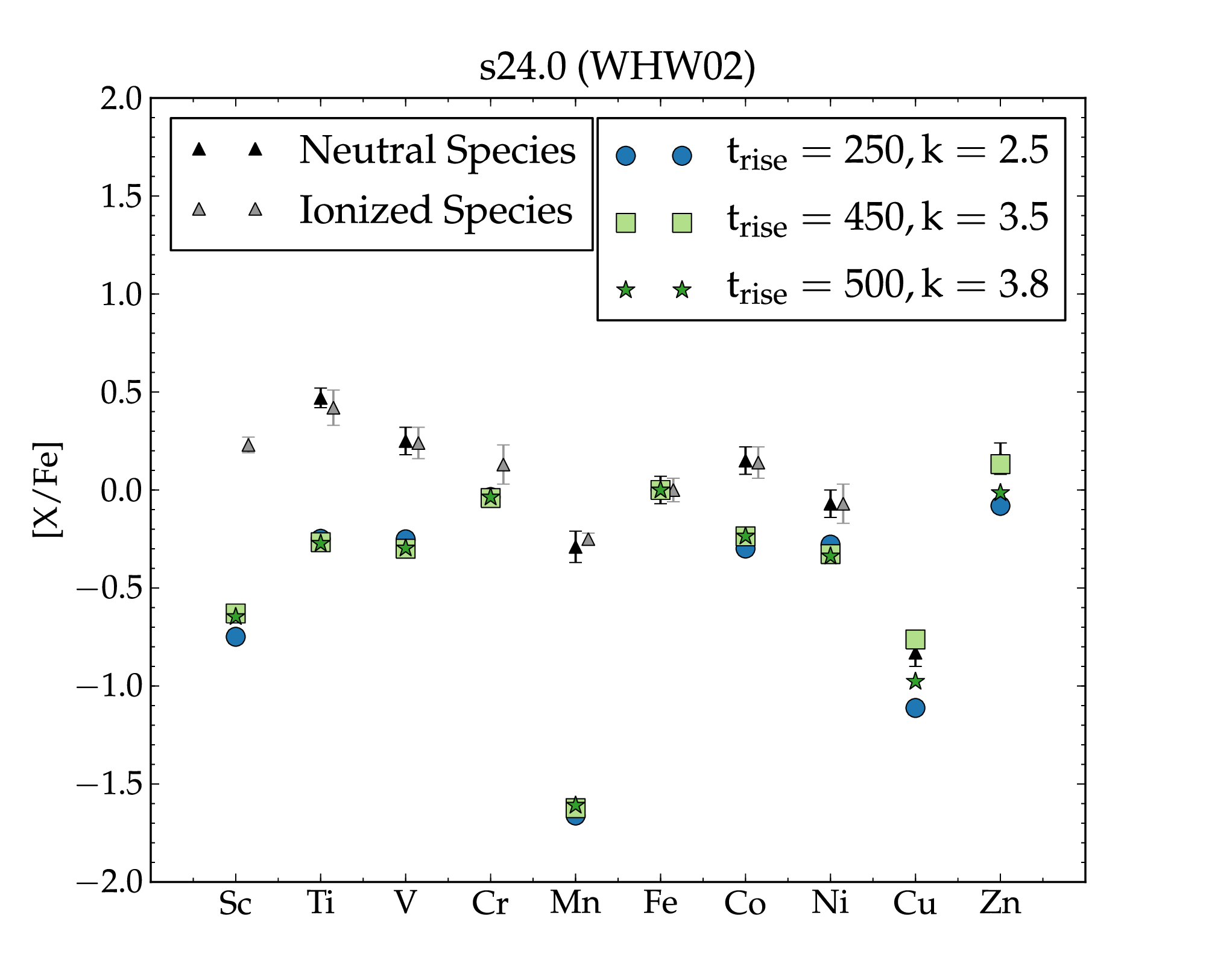} & \includegraphics[width=0.48\textwidth]{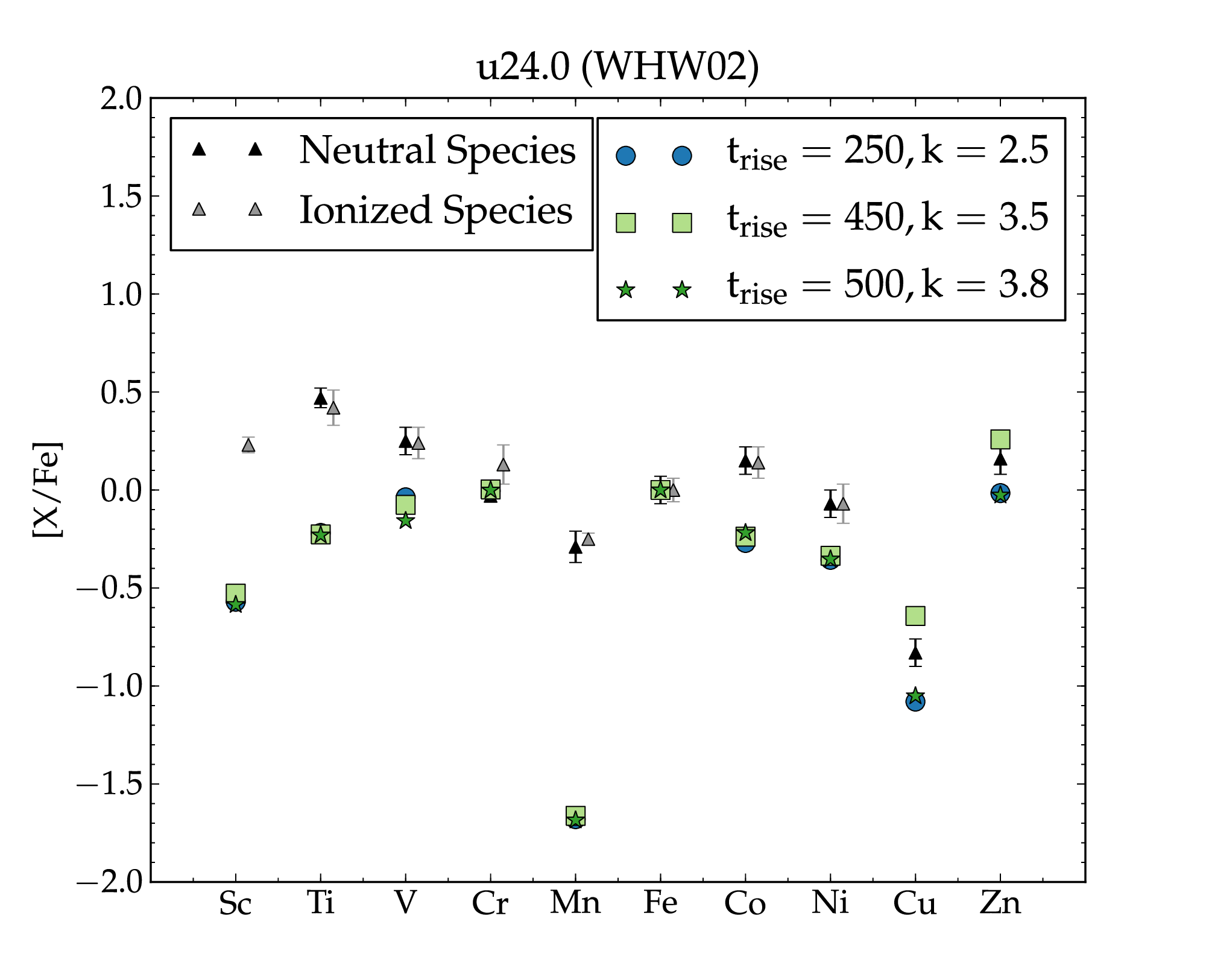} \\
\end{tabular}
\caption{Observed abundances of HD 84937 compared with the yields for solar (left panel) and sub-solar (right panel) metallicity 24M${_\odot}$ WHW02 progenitors.}
\label{tab:s24andu24}
\end{figure}

Sneden et al. (2016) \cite{sneden} used the most recent and improved laboratory data for Fe-group neutral and singly-ionized transitions to derive robust abundances in the very metal-poor main sequence turnoff star HD 84937. Figure \ref{tab:s24andu24} shows the reported abundance ratios of Fe-group elements along with the PUSH nucleosynthesis yields for the Fe-group. The yields depicted are for solar and sub-solar metallicity progenitors from Woosley et al. (2002)\cite{whw02}(WHW02). Different combinations of parameters $t_{rise}$ and $k$ correspond to variations in the artificial heating provided by PUSH, selected to satisfy observational constraints from SN1987A. There are no significant variations seen in the [X/Fe] values obtained for the different parameter settings. 

\begin{figure}[tbh]
\centering
\begin{tabular}{ll}
\hspace{-.015\textwidth}\includegraphics[width=0.48\textwidth]{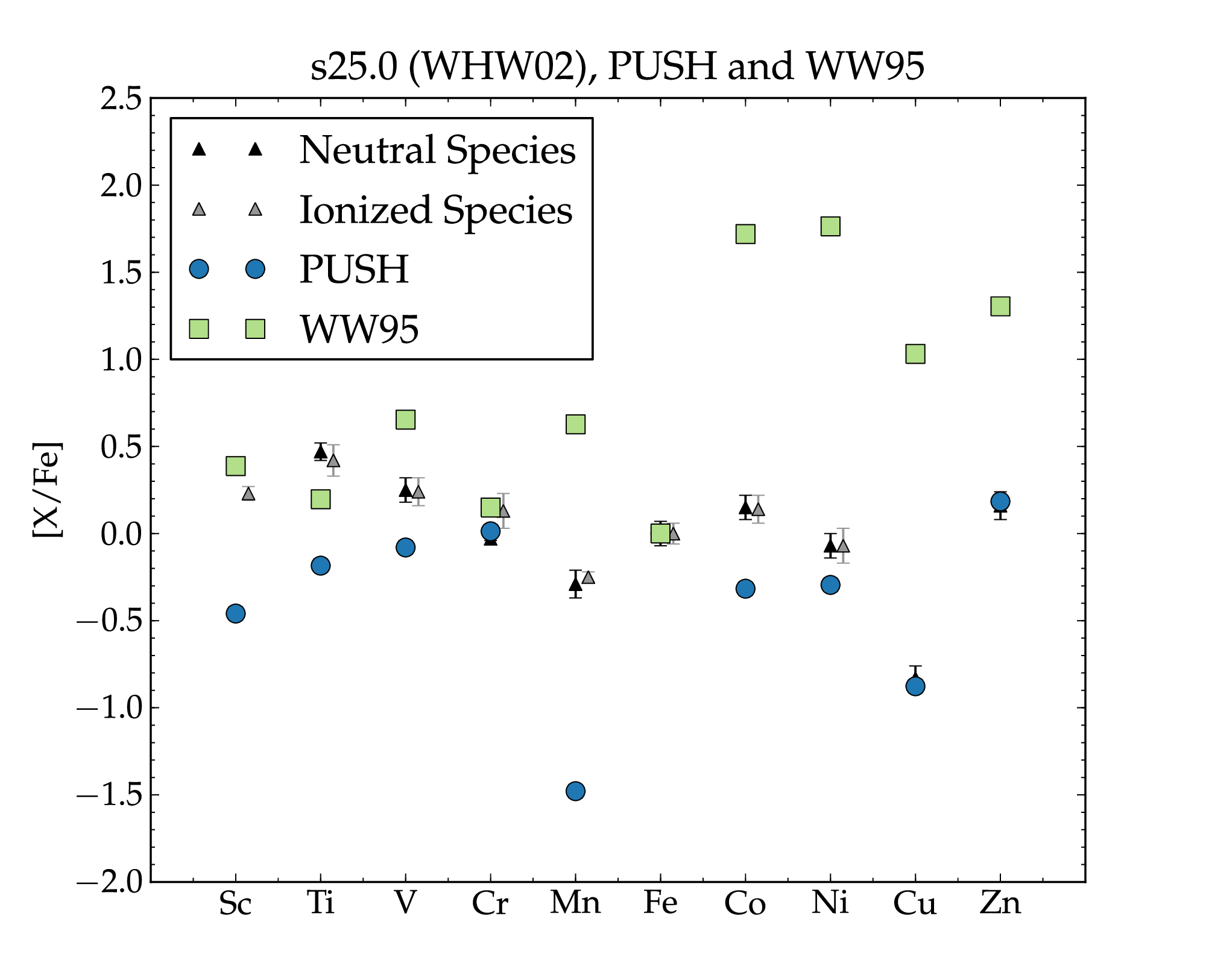} & \includegraphics[width=0.48\textwidth]{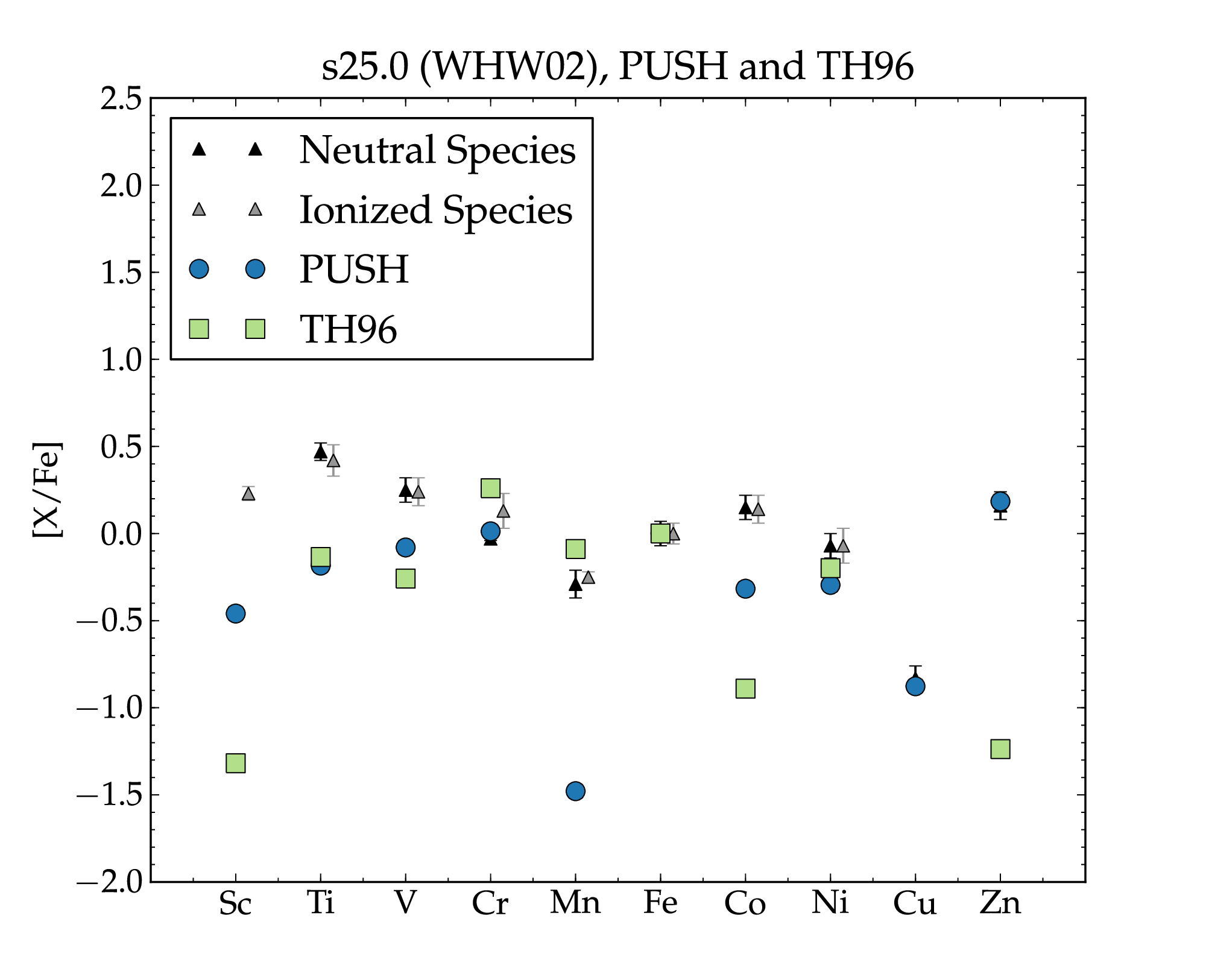} \\
\end{tabular}
\caption{Yields for the solar metallicity 25M${_\odot}$ WHW02 progenitor along with piston (left panel) and thermal bomb (right panel) yields.}
\label{tab:whw02}
\end{figure}

PUSH yields are shown in Figure \ref{tab:whw02} along with piston yields from Woosley et al. (1995) \cite{ww95} and thermal bomb yields from Thielemann et al. (1996) \cite{th96} for a 25M${_\odot}$ model from the WHW02 progenitor set. For most members of this progenitor set, Manganese ratios are found to be very low compared to those seen in HD 84937.

\begin{figure}[tbh]
\centering
\begin{tabular}{ll}
\hspace{-.015\textwidth}\includegraphics[width=0.48\textwidth]{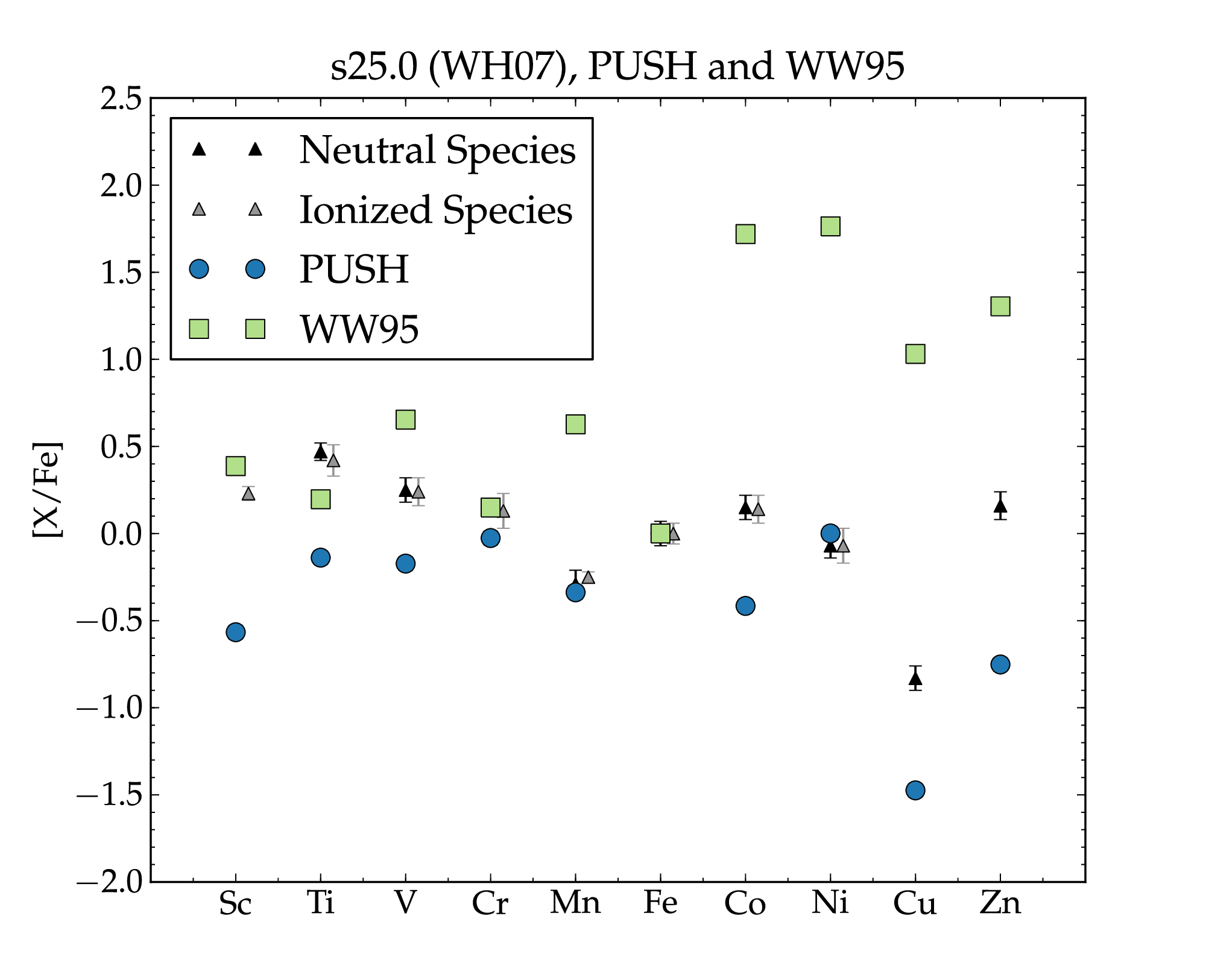} & \includegraphics[width=0.48\textwidth]{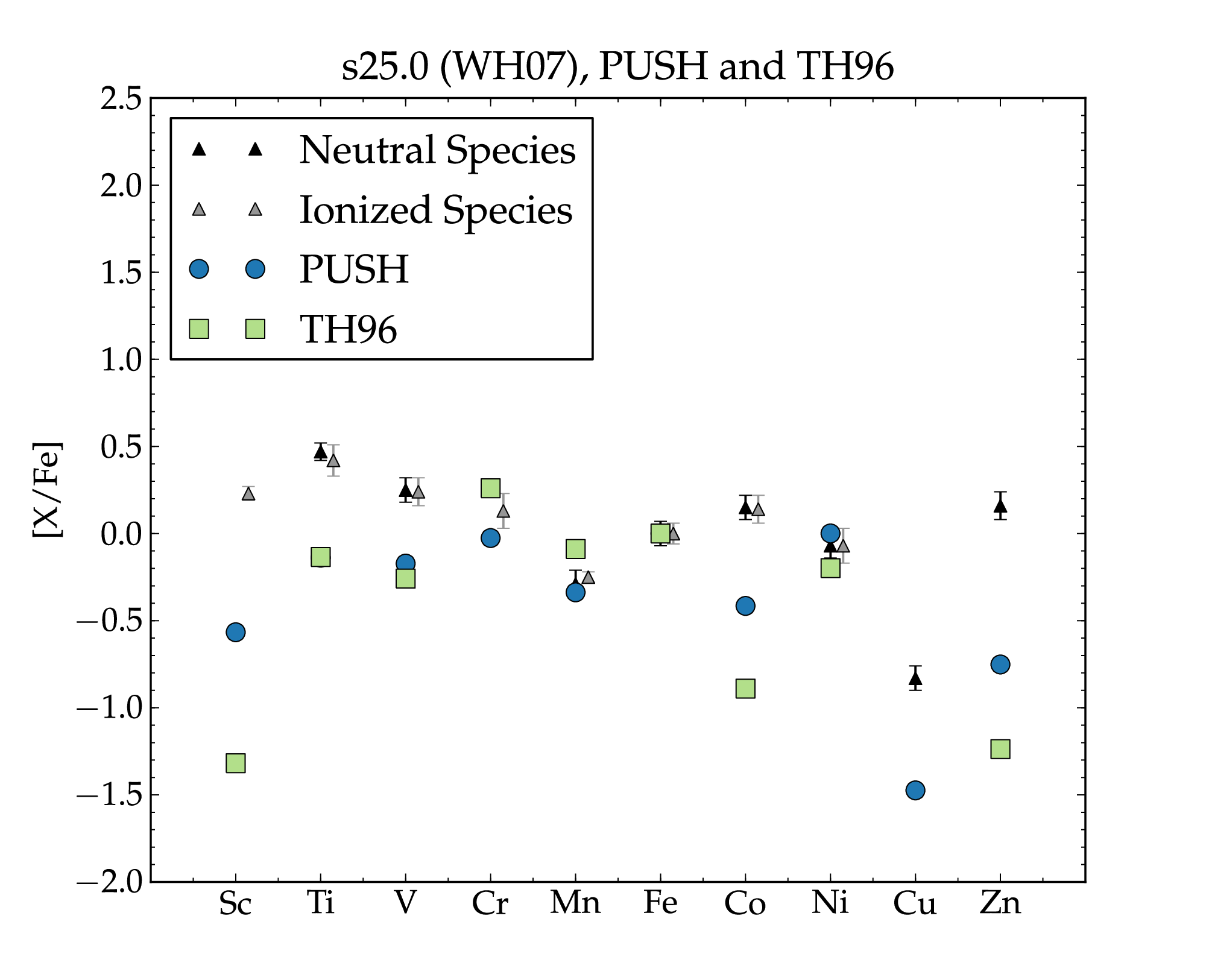} \\
\end{tabular}
\caption{Yields for the solar metallicity 25M${_\odot}$ WH07 progenitor along with piston (left panel) and thermal bomb (right panel) yields.}
\label{tab:wh07}
\end{figure}

PUSH yields were calculated for the Woosley et al. (2007) \cite{wh07} (WH07) progenitor series in addition to WHW02 progenitors. Most of the Fe-group yields for WH07 progenitors are similar to those for WHW02 progenitors but Manganese ratios show a marked improvement. Figure \ref{tab:wh07} shows the calculated yields for a 25M${_\odot}$ WH07 progenitor.

\section{Conclusions and Outlook}
We find that explosions with detailed $Y_e$ evolution give a much better match with observational abundances in HD 84937. Additionally, the Fe-group nucleosynthesis yields appear to depend significantly on the choice of progenitor series.

Nucleosynthesis yields will be calculated for all elements and over a wide range of progenitor masses in a future work \cite{me}, available as input for models of galactic chemical evolution.

\section*{Acknowledgments}
This work is supported through an Early Career Award by the United States Department of Energy (DOE grant no. SC0010263).

\end{document}